\documentclass{article}

\usepackage{amsmath, amsthm, amssymb,color,xcolor,fullpage,url,booktabs}  %cite
\usepackage{graphicx}
\ifdefined\blackandwhite
\usepackage[pdftex]{hyperref} %pagebackref
\else
\usepackage[colorlinks,pdftex]{hyperref} %pagebackref
\hypersetup{citecolor = blue, linkcolor = red!70!black}
\fi

\newcommand{\nc}{\newcommand}
\nc{\rnc}{\renewcommand}

\newcommand{\ket}[1]{\left|#1\right\rangle}

\DeclareMathOperator{\poly}{poly}

\DeclareMathOperator{\BQP}{\mathsf{BQP}}

\DeclareMathOperator{\NP}{\mathsf{NP}}

\DeclareMathOperator{\PH}{\mathsf{PH}}

\DeclareMathOperator{\Ptime}{\mathsf{P}}

\DeclareMathOperator{\PostBPP}{\mathsf{PostBPP}}
\DeclareMathOperator{\PostBQP}{\mathsf{PostBQP}}

\DeclareMathOperator{\PP}{\mathsf{PP}}

\def\be#1\ee{\begin{equation}#1\end{equation}}
\def\bea#1\eea{\begin{eqnarray}#1\end{eqnarray}}
\def\beas#1\eeas{\begin{eqnarray*}#1\end{eqnarray*}}
\def\ba#1\ea{\begin{align}#1\end{align}}
\def\bas#1\eas{\begin{align*}#1\end{align*}}
\def\bpm#1\epm{\begin{pmatrix}#1\end{pmatrix}}
\nc{\non}{\nonumber}
\nc{\nn}{\nonumber}
\nc{\eq}[1]{(\ref{eq:#1})}
\nc{\eqs}[2]{(\ref{eq:#1}) and (\ref{eq:#2})}
\rnc{\L}{\left} 
\nc{\R}{\right}
\nc{\ra}{\rightarrow}
\nc{\ot}{\otimes}
\nc{\grad}{{\vec{\nabla}}}

\newtheorem{thm}{Theorem}
\newtheorem*{thm*}{Theorem}

\newtheorem{proto}{Protocol}

\theoremstyle{definition}

\newtheorem{dfn}[thm]{Definition}
\theoremstyle{plain}

\makeatletter
\newtheorem*{rep@theorem}{\rep@title}
\newcommand{\newreptheorem}[2]{%
\newenvironment{rep#1}[1]{%
 \def\rep@title{#2 \ref{##1} (restatement)}%
 \begin{rep@theorem}}%
 {\end{rep@theorem}}}
\makeatother

\newreptheorem{thm}{Theorem}
\newreptheorem{lem}{Lemma}

\nc\eps{\epsilon}

\nc\cA{\mathcal{A}}
\nc\cB{\mathcal{B}}
\nc\cC{\mathcal{C}}
\nc\cD{\mathcal{D}}
\nc\cE{\mathcal{E}}
\nc\cF{\mathcal{F}}
\nc\cG{\mathcal{G}}
\nc\cH{\mathcal{H}}
\nc\cI{\mathcal{I}}
\nc\cJ{\mathcal{J}}
\nc\cK{\mathcal{K}}
\nc\cL{\mathcal{L}}
\nc\cM{\mathcal{M}}
\nc\cN{\mathcal{N}}
\nc\cO{\mathcal{O}}
\nc\cP{\mathcal{P}}
\nc\cQ{\mathcal{Q}}
\nc\cR{\mathcal{R}}
\nc\cS{\mathcal{S}}
\nc\cT{\mathcal{T}}
\nc\cU{\mathcal{U}}
\nc\cV{\mathcal{V}}
\nc\cW{\mathcal{W}}
\nc\cX{\mathcal{X}}
\nc\cY{\mathcal{Y}}
\nc\cZ{\mathcal{Z}}

\nc\bbC{\mathbb{C}}

\nc\bbF{\mathbb{F}}
\nc\bbM{\mathbb{M}}
\nc\bbN{\mathbb{N}}
\nc\bbR{\mathbb{R}}
\nc\bbZ{\mathbb{Z}}

\nc\benum{\begin{enumerate}}
\nc\eenum{\end{enumerate}}
\nc\bit{\begin{itemize}}
\nc\eit{\end{itemize}}
\newcommand{\fig}[1]{Fig.~\ref{fig:#1}}

\nc{\todo}[1]{\textcolor{red}{todo: #1}}
\nc{\Hnote}[1]{\textcolor{brown}{Aram note: #1}}
\nc{\Mnote}[1]{\textcolor{orange}{Ashley note: #1}}

\def\begsub#1#2\endsub{\begin{subequations}\label{eq:#1}\begin{align}#2\end{align}\end{subequations}}
\nc\qand{\qquad\text{and}\qquad}
\nc\mnb[1]{\medskip\noindent{\bf #1}}

\nc{\pder}[2]{\frac{\partial {#1}}{\partial {#2}}}
\nc{\p}{\partial}

\usepackage{xy}
\xyoption{matrix}
\xyoption{frame}
\xyoption{arrow}
\xyoption{arc}

\usepackage{ifpdf}
\ifpdf
\else
\PackageWarningNoLine{Qcircuit}{Qcircuit is loading in Postscript mode.  The Xy-pic options ps and dvips will be loaded.  If you wish to use other Postscript drivers for Xy-pic, you must modify the code in Qcircuit.tex}
\xyoption{ps}
\xyoption{dvips}
\fi

\entrymodifiers={!C\entrybox}

\renewcommand{\ket}[1]{{\vert{#1}\rangle}}
\newcommand{\qw}[1][-1]{\ar @{-} [0,#1]}
\newcommand{\qwx}[1][-1]{\ar @{-} [#1,0]}
\newcommand{\gate}[1]{*+<.6em>{#1} \POS ="i","i"+UR;"i"+UL **\dir{-};"i"+DL **\dir{-};"i"+DR **\dir{-};"i"+UR **\dir{-},"i" \qw}
\newcommand{\meter}{*=<1.8em,1.4em>{\xy ="j","j"-<.778em,.322em>;{"j"+<.778em,-.322em> \ellipse ur,_{}},"j"-<0em,.4em>;p+<.5em,.9em> **\dir{-},"j"+<2.2em,2.2em>*{},"j"-<2.2em,2.2em>*{} \endxy} \POS ="i","i"+UR;"i"+UL **\dir{-};"i"+DL **\dir{-};"i"+DR **\dir{-};"i"+UR **\dir{-},"i" \qw}
\newcommand{\control}{*!<0em,.025em>-=-<.2em>{\bullet}}
\newcommand{\ctrl}[1]{\control \qwx[#1] \qw}
\newcommand{\lstick}[1]{*!R!<.5em,0em>=<0em>{#1}}
\newcommand{\Qcircuit}{\xymatrix @*=<0em>}
\usepackage{wrapfig}
\usepackage{tcolorbox}
\usepackage{tikz}
\usepackage{multirow}
\usepackage[superscript]{cite}
\nc{\scite}{\cite}

\begin{document}
\title{Quantum Computational Supremacy} % and Computing Beyond the Classical Frontier}
\author{Aram W. Harrow\thanks{Center for Theoretical Physics, MIT. aram@mit.edu}
 \and Ashley Montanaro\thanks{School of Mathematics, University of Bristol, UK.  ashley.montanaro@bristol.ac.uk}}
\maketitle

%\begin{abstract}
%  The goal of quantum computing has always been to do things that cannot be achieved by classical computers, a milestone which has been recently described as ``quantum supremacy''.  The modern approach to quantum supremacy takes it seriously as a goal in itself, which means describing a clear task that can be performed by a realistic near-term quantum computer but not by any classical computer, at least according to some plausible complexity-theoretic assumptions.  This is a departure from the old approach of just aiming to implement a challenging quantum algorithm in that it may involve tasks that are useless in themselves but can convincingly demonstrate advantage over classical computers.  Strengths of this approach include making an explicit comparison with classical competitors and being able to exploit models of quantum computing that are too weak for universal computing.  Weaknesses include the difficulties of verification and the fact that in the long run, supremacy will not be the main selling point of quantum computing.  In this article, we will review the  theoretical and experimental work to date on quantum supremacy, contrast it with the old approach of just aiming to implement a challenging quantum algorithm, and give our perspective on possible future work.
%  \end{abstract}

{\bf The original motivation for quantum computing was the apparent exponential overhead in simulating quantum mechanics on a classical computer.  Following this idea led to the field of quantum algorithms, which aims to find quantum speedups for useful problems, some without obvious relation to quantum mechanics.  A key milestone in this field will be when a universal quantum computer performs a computational task that is beyond the capability of any classical computer, an event known as {\em quantum supremacy}.  This would be easier to achieve experimentally than full-scale quantum computing but  involves new theoretical challenges.}

As a goal, quantum supremacy\scite{preskill12} is unlike most algorithmic tasks since it is defined not in terms of a particular problem to be solved but in terms of what classical computers {\em cannot} do.  This is like the situation in cryptography, where the goal is not only for the authorized parties to perform some task, but to do so in a way that restricts the capabilities of unauthorized parties.
Understanding the fundamental limitations of computation is the remit of the theory of computational complexity~\scite{papadimitriou94}. A basic goal of this theory is to classify problems (such as integer factorisation) into complexity classes (such as the famous classes $\Ptime$ and $\NP$), and then to rigorously prove that these classes are unequal.
In both the case of cryptography and quantum supremacy, computational complexity theory is extremely far from being able to unconditionally prove the desired no-go theorems.  Just as we cannot yet prove that $\Ptime\neq\NP$, we currently cannot unconditionally prove that quantum mechanics cannot be simulated classically.  Instead claims of quantum supremacy will need to rely on complexity-theoretic assumptions, which in turn can be justified heuristically.

\section*{Requirements for quantum supremacy}

Any proposal for a quantum supremacy experiment must have the following ingredients: (1) a well-defined computational task, (2) a plausible quantum algorithm for the problem, (3) an amount of time/space allowed to any classical competitor, (4) a complexity-theoretic assumption (as we will discuss below), and optionally (5) a verification method that can efficiently distinguish the quantum algorithm from any classical competitor using the allowed resources.  Here ``plausible'' means ideally on near-term hardware and will likely include the need to handle noise and experimental imperfections.
We will briefly describe some of the leading approaches to quantum supremacy in these terms.

Note that we do not require that the computational task is of practical interest. When discussing quantum supremacy, it is natural to ask what this gives us that other quantum algorithms, such as factoring~\scite{Shor94} or quantum simulation~\scite{georgescu14,cirac12}, do not.  Indeed, both could be said to be routes to quantum supremacy in their own right.  

For factoring the computational assumption is simply that classical computers cannot factor quickly (say, faster than the current best known algorithm) and the successful operation of a quantum factoring device could be easily verified.  However, the best current estimates suggest that the quantum algorithm requires $\approx 4000$ qubits and $\approx 10^9$ gates\scite{HRS16} to factor a 2048-bit number, and if overheads from fault-tolerance or architectural restrictions are added they could further raise the qubit and gate costs significantly.

Analog quantum simulators\scite{georgescu14,cirac12,Cheuk16}, on the other hand, are already being used to estimate properties of quantum systems which we do not know how to efficiently calculate classically.  If we believe that these properties {\em cannot} be calculated classically, then these experiments could already be considered demonstrations of quantum supremacy.  However, our confidence in conjectures such as ``The correlation functions of a strongly coupled Fermi gas are hard to calculate'' is much lower than our confidence in the hardness of factoring or (as we will discuss below) the non-collapse of the Polynomial Hierarchy.    We will discuss this point in more detail in the subsequent section on complexity theory.

Modern supremacy proposals include constant-depth circuits\scite{TD04}, single photons passing through a linear-optical network (aka ``boson sampling'')\scite{Aaronson13} and random quantum circuits containing gates which either all commute\scite{SB09,BJS10} (a model known as ``IQP'') or do not commute\scite{google-circuit16}. These latter two are described in Boxes 1 and 2.
 In each of these cases, we will describe below arguments why classical simulation is hard even though these models are believed not to be capable of universal quantum computing.
  These occupy a sweet spot between factoring and analog simulation: they can be implemented with much less effort than factoring, including using a non-universal architecture, while the  complexity-theoretic evidence for their superiority over classical computing is stronger than the evidence in favor of specific simulations.  In the sections below we will describe the arguments from complexity theory, and discuss the complications that arise from experimental imperfections and the problem of verification. We summarize some of the main proposals for quantum supremacy in Table~\ref{tab:approaches}.

%TC:ignore
\begin{table}
%\vspace{1em}
\begin{center}
\begin{tabular}{|c|c|c|c|c|}
\hline
\multirow{2}{*}{\bf Algorithm} &
{\bf Difficulty for}
%& \parbox[t]{26ex}{\bf Assumption implying\\ no classical simulation}
& {\bf Assumption implying}
%& \parbox[t]{8ex}{\bf Easy to verify?}
& {\bf Easy to}
& \multirow{2}{*}{\bf Useful?} \\
& {\bf quantum computers} & {\bf no classical simulation} & {\bf verify?} & \\
\hline 
factoring\scite{Shor94} & hard & RSA secure & yes & yes \\ \hline
boson sampling\scite{Aaronson13} & easy & PH infinite 
%\multirow{4}{*}{PH infinite\\ or exact counting $\neq$ approx counting} 
& no & no \\ \cline{1-2} \cline{4-5}
low-depth circuits\scite{TD04} & moderate & or & no* & no \\ \cline{1-2} \cline{4-5}
IQP\scite{SB09} & moderate & approx.~counting & sometimes & no \\ \cline{1-2} \cline{4-5}
QAOA\scite{FarhiH16} & moderate & $\neq$ exact counting& no* & maybe \\ \hline
random circuits\scite{google-circuit16} & moderate & QUATH~(see \cite{aaronson16}) & no & no \\ \hline
adiabatic & \multirow{2}{*}{easy} & \multirow{2}{*}{unknown} & \multirow{2}{*}{no*} & \multirow{2}{*}{maybe} \\
optimization\scite{farhi00} & & & &\\ \hline
analog simulation\scite{cirac12,georgescu14} & easy & idiosyncratic & no & often \\ \hline
\end{tabular}
\end{center}
\caption{{\bf Approaches to quantum supremacy.} Boson sampling, IQP and random circuits are discussed in Boxes 2 and 3. Low-depth circuits are quantum circuits on many qubits, but with only a few layers of quantum gates. QAOA (``Quantum Approximate Optimization Algorithm'') and adiabatic optimization are quantum algorithms for finding reasonably good solutions to optimization problems. Analog simulation is the engineering of one quantum Hamiltonian to directly reproduce the behaviour of another.
The ``difficulty'' column can be viewed as a very crude estimate of how far we would need to proceed towards building a universal quantum computer in order to carry out each algorithm.
For verification we write ``no*'' to mean that we cannot fully verify the validity of the algorithm but we can check some properties such as few-qubit statistics, or the value of the objective function.  We note that outputs of IQP circuits cannot be verified in general, but there is an IQP-based protocol for hiding a string in a code which does have an efficient verification procedure\scite{SB09}.  \label{tab:approaches}}
\end{table}

\section*{Specific proposals for quantum supremacy}

%\begin{tcolorbox}

\begin{figure}
\begin{center}
\begin{tikzpicture}[xscale=0.5,yscale=0.8]
\foreach \x in {0,1,2,3}{
\begin{scope}[xshift=6*\x cm]
\draw[thick] plot[smooth] coordinates { (0.5,0) (1,0.05) (1.5,0.5) (2.5,0.5) (3,0.05) (3.5,0) };
\begin{scope}[yscale=-1,xshift=3cm]
%\draw[thick] plot[smooth] coordinates { (0.5,0) (1,0.05) (1.5,0.5) (2.5,0.5) (3,0.05) (3.5,0) };
\end{scope}
\end{scope}
}
\foreach \x in {0,1,2}{
\begin{scope}[xshift=6*\x cm]
\draw[thick] (3.5,0) -- (6.5,0);
\end{scope}
}
\draw[thick] (0.5,2.4) -- (3.5,2.4);
\draw[thick] (18.5,2.4) -- (21.5,2.4);
\begin{scope}[xshift=0cm,yshift=1.2cm,yscale=-1]
\draw[thick] plot[smooth] coordinates { (0.5,0) (1,0.05) (1.5,0.5) (2.5,0.5) (3,0.05) (3.5,0) };
\end{scope}
\foreach \x in {0.5,1.5,2.5}{
\begin{scope}[xshift=6*\x cm,yshift=1.2cm]
\draw[thick] plot[smooth] coordinates { (0.5,0) (1,0.05) (1.5,0.5) (2.5,0.5) (3,0.05) (3.5,0) };
\begin{scope}[yscale=-1,xshift=3cm]
\draw[thick] plot[smooth] coordinates { (0.5,0) (1,0.05) (1.5,0.5) (2.5,0.5) (3,0.05) (3.5,0) };
\end{scope}
\end{scope}
}
\draw[thick] (0.5,3.6) -- (6.5,3.6);
\draw[thick] (15.5,3.6) -- (21.5,3.6);
\begin{scope}[xshift=3cm,yshift=2.4cm,yscale=-1]
\draw[thick] plot[smooth] coordinates { (0.5,0) (1,0.05) (1.5,0.5) (2.5,0.5) (3,0.05) (3.5,0) };
\end{scope}
\foreach \x in {1,2}{
\begin{scope}[xshift=6*\x cm,yshift=2.4cm]
\draw[thick] plot[smooth] coordinates { (0.5,0) (1,0.05) (1.5,0.5) (2.5,0.5) (3,0.05) (3.5,0) };
\begin{scope}[yscale=-1,xshift=3cm]
\draw[thick] plot[smooth] coordinates { (0.5,0) (1,0.05) (1.5,0.5) (2.5,0.5) (3,0.05) (3.5,0) };
\end{scope}
\end{scope}
}
\draw[thick] (0.5,4.8) -- (9.5,4.8);
\draw[thick] (12.5,4.8) -- (21.5,4.8);
\begin{scope}[xshift=6cm,yshift=3.6cm,yscale=-1]
\draw[thick] plot[smooth] coordinates { (0.5,0) (1,0.05) (1.5,0.5) (2.5,0.5) (3,0.05) (3.5,0) };
\end{scope}
\begin{scope}[xshift=9cm,yshift=3.6cm]
\draw[thick] plot[smooth] coordinates { (0.5,0) (1,0.05) (1.5,0.5) (2.5,0.5) (3,0.05) (3.5,0) };
\begin{scope}[yscale=-1,xshift=3cm]
\draw[thick] plot[smooth] coordinates { (0.5,0) (1,0.05) (1.5,0.5) (2.5,0.5) (3,0.05) (3.5,0) };
\end{scope}
\end{scope}
\begin{scope}[xshift=9cm,yshift=4.8cm,yscale=-1]
\draw[thick] plot[smooth] coordinates { (0.5,0) (1,0.05) (1.5,0.5) (2.5,0.5) (3,0.05) (3.5,0) };
\end{scope}
\begin{scope}[xshift=-1cm,yshift=3.6cm]
\begin{scope}[scale=0.3]
\draw[red,thick] (0,-1) sin (1,1) cos (2,0) sin (3,-1) cos (4,1);
\end{scope}
\end{scope}
\begin{scope}[xshift=-1cm,yshift=2.4cm]
\begin{scope}[scale=0.3]
\draw[red,thick] (0,-1) sin (1,1) cos (2,0) sin (3,-1) cos (4,1);
\end{scope}
\end{scope}
\begin{scope}[xshift=-1cm,yshift=0cm]
\begin{scope}[scale=0.3]
\draw[red,thick] (0,-1) sin (1,1) cos (2,0) sin (3,-1) cos (4,1);
\end{scope}
\end{scope}
\begin{scope}[xshift=22cm,yshift=3.6cm]
\begin{scope}[scale=0.3]
\draw[red,thick] (0,-1) sin (1,1) cos (2,0) sin (3,-1) cos (4,1);
\end{scope}
\end{scope}
\begin{scope}[xshift=22cm,yshift=4.8cm]
\begin{scope}[scale=0.3]
\draw[red,thick] (0,-1) sin (1,1) cos (2,0) sin (3,-1) cos (4,1);
\end{scope}
\end{scope}
\begin{scope}[xshift=22cm,yshift=0cm]
\begin{scope}[scale=0.3]
\draw[red,thick] (0,-1) sin (1,1) cos (2,0) sin (3,-1) cos (4,1);
\end{scope}
\end{scope}
\end{tikzpicture}
\end{center}
\caption{Schematic of a boson sampling experiment. Photons are injected (on left-hand side) into a network of beamsplitters that are set up to generate a random unitary transformation. They are detected on the right-hand side according to a probability distribution conjectured to be hard to sample from classically.\label{fig:bs}}
\end{figure}
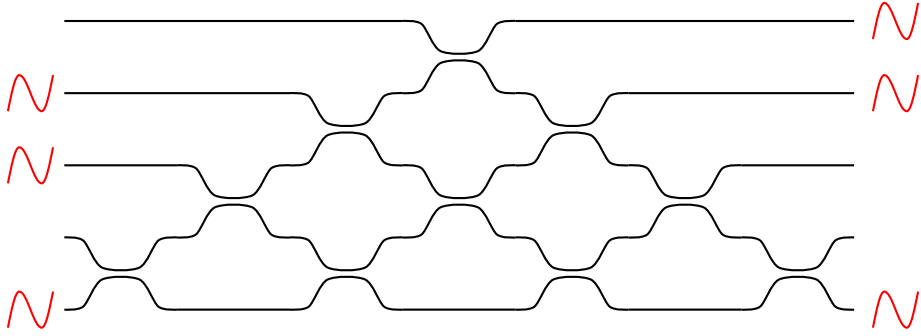

\paragraph{Boson sampling}~\\ 
Boson sampling\scite{Aaronson13} is a formalisation of the problem of simulating noninteracting photons in linear optics; see Figure \ref{fig:bs}. $n$ coincident photons are input into a linear-optical network on $m \gg n$ modes (usually generated at random), with detectors positioned at the output of the network. The challenge is to sample from the distribution on detection outcomes. Following the initial theoretical proposal of Aaronson and Arkhipov\scite{Aaronson13}, several experimental groups quickly demonstrated small-scale examples of boson sampling experiments, with up to 4 coincident photons in up to 6 modes\scite{broome13,spring13,tillmann13,crespi13}. Subsequent work has experimentally validated boson sampling, in the sense of implementing statistical tests that distinguish the boson sampling distribution from other particular distributions\scite{spagnolo14,carolan14}. The current records for implementation of arbitrary linear-optical transformations are 6 modes with up to 6 photons\scite{carolan15} or 9 modes with up to 5 photons\scite{spagnolo14,carolan14,Wang2016}.

Initial boson-sampling experiments used single-photon sources based on spontaneous parametric downconversion. This is a randomised process which has inherently poor scaling with the number of photons, requiring exponential time in the number of photons for each valid experimental run. A variant of boson sampling known as ``scattershot'' boson sampling has therefore been proposed. This uses many sources, each of which produces a photon with some small probability, and it is known in which modes a photon has been produced. Scattershot boson sampling has been implemented with 6 sources and 13 modes\scite{bentivegna15}. An alternative approach is to use a high-performance quantum dot source\scite{Wang2016}.

Challenges faced by experimental implementations of boson sampling include handling realistic levels of loss in the network, and the possibility of the development of more efficient classical sampling techniques.
%\end{tcolorbox}

\begin{figure}[h]
\[
\Qcircuit @C=1em @R=.7em {
\lstick{\ket{0}} & \gate{H} & \ctrl{1} & \ctrl{2} & \qw & \ctrl{3} & \qw & \gate{T^7} & \gate{H} & \meter  \\
\lstick{\ket{0}} & \gate{H} & \gate{Z^{\frac{1}{2}}} & \qw & \gate{Z^{\frac{3}{2}}} & \qw & \ctrl{1} & \gate{T^4} & \gate{H} & \meter  \\
\lstick{\ket{0}} & \gate{H} & \gate{Z^{\frac{1}{2}}} & \gate{Z} & \qw & \qw & \gate{Z} & \qw & \gate{H} & \meter \\
\lstick{\ket{0}} & \gate{H} & \ctrl{-1} & \qw & \ctrl{-2} & \gate{Z} & \qw & \gate{T} & \gate{H} & \meter \\
}
\]
\caption{Example of an IQP circuit. Between two layers of Hadamard gates is a collection of diagonal gates. Although these diagonal gates may act on the same qubit many times, they all commute so in principle could be applied simultaneously.\label{fig:iqp}}
\end{figure}
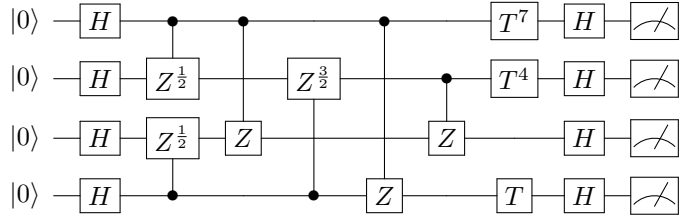

%\begin{tcolorbox}
\paragraph{Random quantum circuits}~\\ 
Unlike boson sampling, some quantum supremacy proposals remain within the standard quantum circuit model. In the model of commuting quantum circuits\scite{SB09}, known as IQP (for ``Instantaneous Quantum Polynomial-time''), one considers circuits made up of gates which all commute, and in particular are all diagonal in the X basis; see Figure \ref{fig:iqp}. Although these diagonal gates may act on the same qubit many times, as they all commute, in principle they could be applied simultaneously. The computational task is to sample from the distribution on measurement outcomes for a random circuit of this form, given a fixed input state. Such circuits are both potentially easier to implement than general quantum circuits, and have appealing theoretical properties which make them simpler to analyse\scite{BJS10,BMS15}. However, this very simplicity may make them easier to simulate classically too.

\begin{figure}[h]
\begin{center}
\begin{tikzpicture}[very thick,scale=0.7]
\draw[blue] (2,1) -- (3,1); \draw[blue] (2,3) -- (3,3); \draw[blue] (2,5) -- (3,5);
\draw[blue] (0,0) -- (1,0); \draw[blue] (0,2) -- (1,2); \draw[blue] (0,4) -- (1,4);
\draw[blue] (4,0) -- (5,0); \draw[blue] (4,2) -- (5,2); \draw[blue] (4,4) -- (5,4);
\foreach \x in {0,...,5}{
\foreach \y in {0,...,5}{
\fill (\x,\y) circle (0.1);
}}
\foreach \x in {0,1,4,5}{
\foreach \y in {1,3,5}{
\fill[red] (\x,\y) circle (0.1);
}}
\foreach \y in {0,2,4}{
\fill[red] (2,\y) circle (0.1);
\fill[red] (3,\y) circle (0.1);
}
\begin{scope}[xshift=15cm]
\begin{scope}[rotate=90]
\draw[blue] (2,1) -- (3,1); \draw[blue] (2,3) -- (3,3); \draw[blue] (2,5) -- (3,5);
\draw[blue] (0,0) -- (1,0); \draw[blue] (0,2) -- (1,2); \draw[blue] (0,4) -- (1,4);
\draw[blue] (4,0) -- (5,0); \draw[blue] (4,2) -- (5,2); \draw[blue] (4,4) -- (5,4);
\foreach \x in {0,...,5}{
\foreach \y in {0,...,5}{
\fill (\x,\y) circle (0.1);
}}
\foreach \x in {0,1,4,5}{
\foreach \y in {1,3,5}{
\fill[red] (\x,\y) circle (0.1);
}}
\foreach \y in {0,2,4}{
\fill[red] (2,\y) circle (0.1);
\fill[red] (3,\y) circle (0.1);
}
\end{scope}
\end{scope}
\end{tikzpicture}
\end{center}
\caption{ The architecture proposed by the quantum-AI group at Google to
  demonstrate quantum supremacy consists of a 2d lattice of
  superconducting qubits.  This figure depicts two illustrative
  timesteps in this proposal.  At each timestep, 2-qubit gates (blue)
  are applied across some pairs of neighbouring qubits, and random
  single-qubit gates (red) are applied on other
  qubits.\label{fig:google}}
\end{figure}
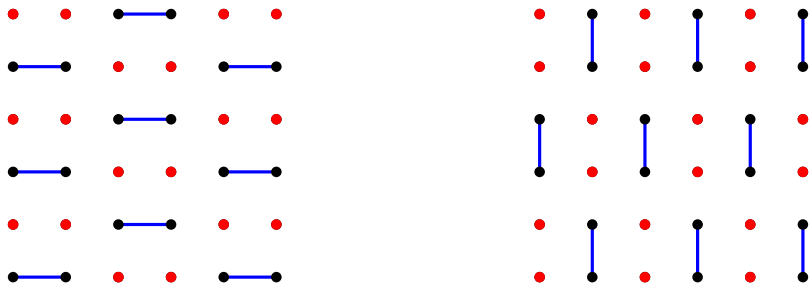

Of course, one need not restrict to commuting circuits to demonstrate supremacy. The quantum-AI group at Google has recently suggested an experiment based on superconducting qubits and noncommuting gates\scite{google-circuit16}. The proposal is to sample from the output distributions of random quantum circuits, of depth around 25, on a system of around 49 qubits arranged in a 2d square lattice structure (see \fig{google}).  It is suggested in \cite{google-circuit16} that this should be hard to simulate, based on (a) the absence of any known simulation requiring less than a petabyte of storage, (b) IQP-style theoretical arguments\scite{BMS15} suggesting that  larger versions of this system should be asymptotically hard to simulate, and (c) numerical evidence\cite{google-circuit16} that such circuits have properties that we would expect in hard-to-simulate distributions.  If this experiment were successful it would come very close to being out of reach of current classical simulation (or validation, for that matter) using current hardware and algorithms.
%\end{tcolorbox}

\section*{Why supremacy?}
Before proceeding, we should discuss why a demonstration of quantum supremacy is worthwhile. The field of  quantum computing is based on the premise that quantum mechanics has changed the definitions of information and computation, with implications that are both philosophical and practical.   For example, entanglement is a useful form of correlation that would not exist in a classical theory of information and its existence can be demonstrated with experiments designed to test Bell-inequality violations.   Supremacy experiments can be thought of as the computational analogue of Bell experiments.  Just as Bell experiments refute Local Hidden Variable models, supremacy experiments refute the old ``Extended Church-Turing (ECT) thesis'', which asserts that classical computers can simulate any physical process with polynomial overhead.   Such a demonstration would be convincing evidence confirming the consensus model of quantum mechanics, showing that the world contains not only entanglement but also is capable of computational feats beyond the reach of classical computers.  Validating the standard picture of quantum mechanics in this way would be valuable for foundational reasons because quantum mechanics is so far the only physical theory to change our model of computing; and for practical reasons because it would greatly increase our confidence in the eventual feasibility of large-scale quantum computing.

The ECT thesis also motivates our focus on quantum mechanics, as opposed to  hard-to-simulate classical systems such as fluid dynamics or protein folding.   With these examples the difficulties are ``merely'' from issues such as separations of scales in time or space, and these in principle could be simulated with effort linear in the energy and space-time volume of the system.  This means that a protein-folding problem which would require $10^{50}$ steps for a naive simulation is not an instance of a family that includes problems requiring $10^{100}$ or $10^{1000}$ steps.  By contrast, a quantum supremacy experiment that barely surpasses our existing classical computers would be significant in part because it would imply that vastly greater separations in computational power are likely to soon follow, as we will explore further in the next section.

\section*{Complexity-theoretic basis for quantum supremacy}
\label{sec:complexity}

Since quantum supremacy is ultimately about comparison between quantum and classical computers, demonstrating it will require some computational assumption about the limits to the power of classical computers.  At a minimum, we need to assume that quantum mechanical systems cannot be simulated efficiently (i.e.~with polynomial overhead) by classical computers.
But just as cryptography always needs assumptions stronger than $\Ptime\neq \NP$, each quantum supremacy proposal needs its own assumption. Although such assumptions must ultimately be at least as strong as the lack of efficient classical simulation of quantum computers, we may hope for them to be based on different principles and to be believable in their own right.

As discussed above, if we use the quantum computer for factoring or simulation, then our assumption should simply be that those problems are hard for classical computers.  Our belief that factoring is hard is based on many mathematician-hours put into solving it; on the other hand, the best known algorithms are only from ca.~1990 and are significantly faster than brute-force search, so further improvements may well exist.

The complexity of quantum simulation is much murkier.  One difference is the great diversity of quantum systems and of methods for treating them, which are often adapted to specific features of the system.  The complexity of a simulation can also vary with parameters such as temperature and coupling strengths in nonobvious ways.
Finally, when analog quantum simulators cannot address individual qubits, this limits their ability to encode a wide range of problem instances, and makes the complexity of the problem they do solve even less clear.  The problems solved by quantum simulators can certainly teach us about physics and often in ways that we do not know how to do classically; however, our confidence that they cannot be classically simulated is rather weak.

We now turn to the modern supremacy proposals. These are often based around sampling problems\scite{lund17} rather than decision problems, where the task is to output samples from a desired distribution, rather than to output a deterministic answer. The strength of these is that, despite working with a restricted model of quantum computing (boson sampling, low-depth circuits, etc.), they do not need to assume that this specific model is hard to simulate.  Indeed the complexity assumption can be expressed in terms of concepts that have been studied since the 1970s and are thought to be hard for reasons that do not rely on any beliefs about quantum mechanics.  One assumption that will work is known as the ``non-collapse of the polynomial hierarchy,'' which we explain in a sidebar.  Another possible assumption is that exact counting of exponentially large sets is harder than approximate counting.  Stronger assumptions are also possible, and these can be used to rule out larger classes of classical simulations or in some cases to enable more efficient verification of the quantum device.

Why are these complexity assumptions relevant to simulating quantum computers?  The main idea is to use a technique called ``post-selection,'' which refers to the following scenario.  A computation, which could be either classical or quantum, takes input string $x$ and outputs strings $y$ and $z$.  The string $y$ is used to represent the output, while we condition (``post-select'') on the string $z$ taking some fixed value, say $00\ldots 0$.  Many experiments post-select in practice (e.g.~on coincident detection events) but usually on events whose probability is not too small.  We will allow post-selection even on exponentially unlikely outcomes, which will make the ability to post-select extremely powerful.  The purpose of post-selection is two-fold.  First, an efficient classical simulation of a quantum computation implies that a classical computer with post-selection can efficiently simulate a quantum computer with post-selection.  However, this latter claim would contradict our assumption that the polynomial hierarchy doesn't collapse, as we will explain in the sidebar.  Second, many non-universal models of quantum computation become universal once post-selection is allowed.   Thus even an efficient classical simulation of one of these restricted models of quantum computing would lead to the same contradictions.

 In a sidebar we describe a somewhat indirect argument which implies that an efficient {\em exact} classical simulation (we discuss approximate simulations below) of any of these restricted models of quantum computing would lead to several surprises, including the collapse of the polynomial hierarchy and exact counting being roughly as hard as approximate counting. Neither of these is believed to be true; conversely, this consensus of complexity theorists implies that there is no efficient exact classical simulation of boson sampling, IQP or the other models.  How strong are these beliefs?  The non-collapse of the polynomial hierarchy is a conjecture that is stronger than $\Ptime\neq\NP$ but that is plausible for similar reasons.  Unlike factoring, there are essentially no non-trivial algorithms which would suggest that the ability to perform approximate counting would yield an exact counting algorithm in anything less than exponential time.  On the other hand, these questions have been studied by a smaller and less diverse group of researchers.   Nevertheless these are conjectures that we should have high confidence in.    Going further, we will describe several stronger (i.e.~less plausible) conjectures that will let us rule out more classical simulations.

%TC:ignore
\begin{tcolorbox}
\paragraph{Sidebar: The Polynomial Hierarchy and Post-Selection}~\\ 

\noindent{\em The polynomial hierarchy.}
To explain the polynomial hierarchy (denoted $\PH$), we start with $\NP$.  A boolean function $f(x)$ is in $\NP$ if it can be expressed as $f(x) = \vee_y g(x,y)$ for some poly-time computable function $g$.  For example, in the 3-coloring problem, $x$ specifies a graph, and $f(x)$ is true if and only if the graph is 3-colorable, i.e.~the vertices can be each colored red, green or blue in a way such that no edge connects two vertices with the same color.  If $y$ is a list of colors for each vertex then we can easily compute $g(x,y)$, which is 1 if each edge connects vertices of different colors, and 0 if not; then $f(x)=\vee_y g(x,y)$ so the 3-coloring problem is in $\NP$.

The $k^{\text{th}}$ level of  $\PH$ is defined to be the set of functions that can be written in terms of $k$ alternating quantifiers, i.e.~$f(x) = \vee_{y_1} \wedge_{y_2} \vee_{y_3} \cdots \wedge_{y_k} g(x,y_1,\ldots,y_k)$, where $g(\cdot)$ is poly-time computable, and where the length of each $y_1,\ldots,y_k$ grows at most polynomially with the length of $x$.  Just as it is conjectured that $\Ptime \neq \NP$, it is conjectured that the $k^{\text{th}}$ level of the $\PH$ is not equal to the $k+1^{\text{st}}$ level for any $k$.  If this were to be false and there existed some $k$ for which the $k^{\text{th}}$ level equalled the $k+1^{\text{st}}$ level  then the $k^{\text{th}}$ level would also equal the $k+2^{\text{nd}}$ and all later levels as well -- a scenario we refer to as the ``collapse'' of the $\PH$ to the $k^{\text{th}}$ level.  

{\em Post-selection.} Allowing post-selection (described in the main text) dramatically increases the power of both classical and quantum computation.  The corresponding complexity classes are called $\PostBPP$ and $\PostBQP$ respectively.  It turns out that $\PostBPP$ is somewhere between the first and third levels of the $\PH$\scite{han97} and $\PostBQP$ corresponds to the class $\PP$\scite{aaronson05b}, which appears to be much stronger.  Indeed, any problem in $\PH$ can be reduced to solving a polynomial number of problems in $\PP$, or formally $\PH \subseteq \Ptime^{\PP}$\scite{toda91}.  This means that if $\PostBPP$ were to equal $\PostBQP$ then it would imply the collapse the $\PH$.  Conversely, the non-collapse of the $\PH$ implies that post-selected classical and quantum computing are far apart in their computational power.
\end{tcolorbox}

\begin{tcolorbox}
\paragraph{Sidebar: Counting}~\\

Another way to think about the difference between $\PostBPP$ and $\PostBQP$ is in terms of counting problems. Consider a problem in $\NP$ of the form $f(x) = \vee_y g(x,y)$, where again $g$ is poly-time computable, and we represent True and False with 1 and 0 respectively.  Rather than asking whether or not there is a $y$ such that $g(x,y)=1$, we can instead ask how many such $y$ there are.  This corresponds to the function $f_{\text{count}}(x) := \sum_y g(x,y)$.  The class $\NP$ corresponds to determining whether $f_{\text{count}}(x)$ is $=0$ or $\geq 1$.  We can express $\PostBPP$ and $\PostBQP$ as well in terms of $f_{\text{count}}$.  $\PostBQP$ corresponds to being given some threshold $T$ and determining whether $f_{\text{count}}(x)$ is $>T$ or $\leq T$, a task known as {\em exact counting}.  Keep in mind here that $y$ is a string of $\poly(n)$ bits, so that $T$ can be exponentially large in $n$.  By contrast, $\PostBPP$ is equivalent to {\em approximate counting}\scite{odonnell15}: formally, given a threshold $T$ and an accuracy parameter $\eps = 1/\poly(n)$, determine whether $f_{\text{count}}(x)$ is $\geq T(1+\eps)$ or $<T$ given that one of these is the case.

Just as we could start with the assumption that the $\PH$ does not collapse, we could also conjecture that exact counting is much more difficult than approximate counting.  This assumption would equally well imply that $\PostBPP\neq\PostBQP$ and in turn that there does not exist an efficient classical simulation of even restricted models of quantum computing.
\end{tcolorbox}

%TC:endignore

\section*{Fine-grained complexity assumptions}

If we equate ``efficient'' with ``polynomial-time'', then conjecturing that $\PostBPP\neq\PostBQP$ is enough to show that classical computers cannot exactly simulate quantum computers ``efficiently.''  However, these asymptotic statements tell us little about the actual quantum computers being built in the coming years, or about the ability of existing classical competitors to simulate them.  For this, we would like statements of the form ``a quantum circuit with 500 gates on 100 qubits cannot be perfectly simulated using a classical computer with $10^9$ bits of memory using fewer than $10^{12}$ operations.''  Only in this way could we point to the outcomes of a specific experiment and conclude that quantum supremacy had been achieved.  A crucial ingredient here is a concrete or ``fine-grained'' complexity assumption.  
An example of such an assumption is the Exponential Time Hypothesis (ETH)\scite{impagliazzo01} which asserts that 3-SAT instances on $n$ bits require time $\geq 2^{cn}$ (for some constant $c>0$) to solve on a classical computer.  Concrete bounds require an even more explicit hypothesis: for example, we might assert that the ETH holds with $c=0.386$  for random 3-SAT instances with a planted solution, corresponding to the best known algorithm.  These bounds might be based on analysis of the best known algorithm or on provable lower bounds if we replace the 3-SAT instance with a black-box ``oracle'' function.  
Work in progress\scite{DLH17} uses oracle arguments to devise a plausible conjectured variant of ETH which will in turn imply that $\approx 1700$ qubits can carry out a computation which cannot be simulated classically in less than a day by a supercomputer performing $10^{17}$ operations per second.  Improving this is an important open problem, and possible routes to progress include stronger assumptions, better algorithms or a sharper analysis.

\section*{Average-case assumptions}
The best-studied conjectures and results in complexity theory consider {\em worst-case} hardness, meaning that to compute a function one must be able to do so for all possible inputs.  However, in terms of quantum supremacy worst-case hardness assumptions only translate into the statement that there exists {\em some} quantum circuit of a given size which cannot be efficiently simulated, or equivalently that no classical algorithm works for all quantum circuits.  What can an experimentalist do with such statements?  There is no clear guidance on which quantum circuits to implement, and once a circuit has been chosen, no way to certify it as hard.

The same issues arise in cryptography, where we would like to argue that a given cryptosystem is hard to break not merely in the worst case but for {\em most} choices of random key.  Such statements are the domain of {\em average-case complexity}, in which we say that a function $f$ is hard to compute on some distribution $D$ if there is no efficient algorithm whose output equals $f(x)$ for most values of $x$ sampled from $D$.  Average-case hardness conjectures are stronger than (i.e.~less plausible than) worst-case hardness conjectures and cannot be reduced to each other as readily as can the worst-case hardness of NP-complete problems.  Nevertheless there are well-studied distributions of instances of NP-complete problems, such as random 3-SAT\scite{cheeseman91,MMZ06}, where no algorithms are known to run in less than exponential time, and there are distributional versions of NP-completeness\scite{Levin-average}.

The benefits of using average-case assumptions are two-fold.  First, they give us a concrete family of experiments for which we can say that most such experiments are hard to simulate.  For example, we might consider random quantum circuits of a specified size.  Second, and less obviously, they allow us to rule out a larger class of classical simulations\scite{Aaronson13,BMS15,gao16}.   Suppose the quantum circuit outputs string $z$ with probability $q(z)$ and a classical simulator does so with probability $p(z)$.  Worst-case assumptions allow us to argue that for any specific $z$, say $0^n$, it is not possible to find a classical simulator with $p(0^n)=q(0^n)$ for all circuits, or even with low {\em multiplicative} error:
\be 0.9 q(0^n) \leq p(0^n) \leq 1.1 q(0^n).\label{eq:multiplicative}\ee
   While this does indeed rule out highly accurate simulations, we would like to rule out even simulations with what is called low {\em additive} error
\be \sum_z |p(z)-q(z)|\leq 0.001\label{eq:trace-dist}.\ee  
This notion of approximation is natural because, if $p$ and $q$ are close under this distance measure, they cannot be distinguished without taking many samples.
If we make an average-case hardness assumption and prove one more technical assumption known as anticoncentration, we can rule out additive-error simulations (i.e. satisfying \eq{trace-dist}).  (Anticoncentration means that the distribution $q(z)$ is reasonably close to uniform.  It is known to hold for random circuits\scite{aaronson16} and for IQP\scite{BMS15} and is conjectured to hold for boson sampling\scite{Aaronson13}, although it does not hold for constant-depth random circuits~\cite{HM17}.)

One disadvantage of  average-case assumptions is that they cannot easily be reduced to each other.  By contrast, if a problem is NP-hard in the worst case then we know that an algorithm that works for all inputs would yield algorithms for thousands of other problems in NP, which collectively have been studied for decades by researchers across all of science and engineering.  But for average-case hardness, we may have different hardness for each distribution of instances.  For example, for 3-SAT a natural distribution is to choose $n$ variables and $\alpha n$ random clauses.  It is believed that random instances are likely to be satisfiable for $\alpha < \alpha_c$ and unsatisfiable for $\alpha > \alpha_c$, for some critical value $\alpha_c \approx 4.2667$\scite{MMZ06}.   Based on this, it is reasonable to conjecture that choosing $\alpha = \alpha_c$ yields a hard distribution, but this conjecture is far flimsier than the worst-case conjectures even for this relatively well-studied problem.

In rare cases, a problem will have the same average-case and worst-case complexity, and it is a major open question to establish quantum supremacy based on such a problem.  Boson sampling takes steps in that direction\scite{Aaronson13}, by using a conjecture about the average-case complexity of estimating the permanent, while an average-to-worst-case reduction is known only for the exact case.  Indeed the known reduction is based on polynomial interpolation and its numerical instability means that new ideas will be needed to argue that estimating the permanent is hard on average. More generally, a major open problem is to base the hardness of approximate classical simulations of the form of \eq{trace-dist} merely on well-believed classical complexity assumptions, such as non-collapse of the polynomial hierarchy.

\section*{Maximal assumptions}
Another reasonable possibility is to make our complexity assumptions as strong as possible without contradicting known algorithms.  
Here the high-level strategy is to try to improve our (often exponential-time) classical simulations as far as possible and then to conjecture that they are essentially optimal.  Aaronson and Chen\scite{aaronson16} have recently carried out this program.  Among other contributions, they developed classical simulations for $n$-qubit, depth-$d$ circuits that calculate matrix elements in time $O((2d)^n)$ and nearly linear space (note that with $2^n$ space, $O(d2^n)$ time is possible).  An easier task than classical simulation is to distinguish likely from unlikely outcomes of a quantum circuit with some exponentially small advantage over random guessing.  The ``QUATH'' conjecture\scite{aaronson16} asserts that poly-time classical algorithms cannot perform this task for quantum circuits whose depth $d= \Omega(n)$.
The advantage of this approach is that it enables a ``semi-efficient'' verification procedure which uses the quantum device only a polynomial number of times but still requires exponential time on a classical computer.  

Making these conjectures as strong as possible makes our confidence in them as low as possible; essentially any non-trivial improvement in simulating quantum mechanics would refute them. But so what?  Unlike the case of cryptographic assumptions, a too-strong conjecture would not create any vulnerabilities to hackers.  In this view, hardness conjectures are just ways of guessing the complexity of simulating quantum systems, and these estimates are always subject to revision as new evidence (in the form of algorithms) appears.  Further, these conjectures highlight the limits of our current simulation algorithms, so that refuting them would be both plausible and a significant advance in our current knowledge.

\section*{Physical noise and simulation errors}
\label{sec:noise}

Any realistic quantum experiment will be affected by noise, i.e.\ undesired interactions with the environment. Dealing with this noise is a major challenge for both theorists and experimentalists. The general theory of quantum fault-tolerance\scite{knill98b,knill05} allows quantum computations to be protected against a sufficiently small amount of physically reasonable noise. However, although the asymptotic overhead of fault-tolerance is relatively minor, the constant factors involved are daunting: to produce a fault-tolerant logical qubit may require $10^3-10^4$ physical qubits\scite{fowler12}, an overhead far too great for short-term quantum supremacy experiments. As excessive noise can render a hard probability distribution easy to simulate, it is an important question to determine to what extent these experiments remain hard to simulate classically, even in the presence of uncorrected noise.

A related issue is that classical simulation algorithms of quantum circuits will have errors of their own.   This could be seen as analogous to the fact that realistic quantum computers only implement ideal quantum circuits imperfectly.  Classical noise could be multiplicative as in \eq{multiplicative} or additive as in \eq{trace-dist}.  Methods based on representing the exact state~\cite{MS08} can achieve low enough error rates that we can think of them as low multiplicative error, while methods based on sampling (e.g.~\cite{bravyi16a}) naturally achieve low additive error.
For multiplicative noise it is relatively easy to show hardness results.  IQP circuits remain hard to simulate under this notion of noise~\cite{BJS10}, and similar results have since been shown for the one clean qubit model\scite{morimae14} and other restricted classes of circuits.  However, additive noise is arguably a more natural model, and ruling out such simulations would be a stronger result.

Addressing this question was one of the major steps forward taken by Aaronson and Arkhipov\scite{Aaronson13} in their work on boson sampling. Based on two reasonable (yet currently unproven) conjectures, they argued that sampling from the boson sampling distribution should still remain classically hard if the classical sampler is allowed to only approximately sample from the distribution. That is, the classical sampler is asked to output a sample from any distribution whose total variation distance from the true boson sampling distribution is at most a small constant. Assuming their conjectures, as long as the quantum experiment experiences a total amount of noise below this threshold, its output is still hard to sample from classically.

One of the conjectures is a technical claim about anticoncentration of the permanent of a random matrix, with strong numerical evidence for its truth. The other (known as the ``permanent-of-Gaussians'' conjecture) is an average-case hardness assumption asserting that the permanent of a matrix of Gaussian random variables should be hard to approximate up to small relative error. This hardness property can be shown to hold for exact computation of the permanent of such random matrices\scite{Aaronson13}, but extending it to small relative error seems to be beyond the reach of current techniques.

Another step forward was the proof of a similar result for the IQP model\scite{BMS15}. In this case, two conjectures occur which are analogous to those for boson sampling; however, in the setting of IQP the analogue of the anticoncentration conjecture can be proven. The permanent-of-Gaussians conjecture is replaced with equivalent conjectures about either the average-case hardness of approximately computing the partition function of the Ising model, or the average-case hardness of approximately computing a natural property of low-degree polynomials over finite fields\scite{BMS15}. Anticoncentration and average-case hardness conjectures naturally occur in the setting of noisy quantum supremacy experiments because approximating a probability distribution up to small total variation distance is similar to approximating most of the probabilities up to a very small additive error. If most of the probabilities are hard to approximate up to small relative error, and most of them are rather large (i.e.\ the distribution is not too concentrated) then a good classical approximation in total variation distance leads to a contradiction.

The question of how to model simulability in the presence of noise is subtle and still under debate. For example, a counterpoint to these hardness results is provided by recent work showing that, if an {\em arbitrarily small} constant amount of noise occurs on each qubit at the end of an IQP circuit, the class of random IQP circuits which is conjectured hard to simulate\scite{BMS15} can be simulated classically in polynomial time up to small variational-distance  (as in~\eq{trace-dist}) error\scite{bremner16b} . This contrasts with another recent result showing that classical simulation of the noisy distribution up to small {\em relative} error (as in~\eq{multiplicative}) can be hard\scite{fujii16}. As noise at the end of an IQP circuit can be dealt with using simple classical error-correction techniques with low overhead\scite{bremner16b}, this suggests that quantum supremacy experiments may need to make use of some form of error-correction, but this might be substantially simpler than the machinery required for full quantum fault-tolerance.

\section*{Verification}

A key issue for any proposed quantum supremacy experiment is verification of the results of the experiment. In order to claim quantum supremacy, we must have confidence that the experiment has indeed done something which is hard for a classical computer. By definition, quantum supremacy experiments cannot be simulated efficiently classically, so we must seek another means of checking that such an experiment has succeeded. If we had a large-scale quantum computer that could run Shor's algorithm, verification would be easy: we could challenge the experimenter to factor a 2048-bit RSA key, then check that the claimed factors multiplied to the correct number. However, integer factorisation is a rare example of a problem which is both tractable on a quantum computer (in the complexity class $\BQP$\scite{watrous09}), checkable on a classical computer (in the complexity class $\NP$\scite{papadimitriou94}), yet not known to be efficiently solvable on a classical computer. Very few such problems are known, and none are currently known which would be classically intractable for instance sizes small enough to be solved by a quantum computer with, say, 100 logical qubits.

In the short term, then, verification of quantum supremacy  needs to use different methods, none of which is yet as simple and powerful as checking integer factorisation.
 Which approach is preferred may depend on the assumptions one wishes to make about the experiment being performed. This is analogous to the setting of experimental tests of Bell-inequality violations: different techniques can be used to rule out different loopholes, but it is very challenging to rule out all loopholes simultaneously.

One straightforward approach is to build confidence that the experiment (which is hard to test in its entirety) is working correctly by testing smaller parts of it. This could involve testing individual components within a quantum circuit -- a task likely to be required for any experiment anyway -- or running quantum computations which are small or simple enough to be classically simulable. A non-trivial example of this is executing computations which are mostly or entirely comprised of Clifford gates, which are known to be efficiently classically simulable\scite{aaronson04a,bravyi16a} despite displaying such quantum characteristics as allowing the creation of large-scale entanglement. Another example is replacing the random linear-optical transformation used in boson sampling with a highly structured one, such as a quantum Fourier transform\scite{tichy14}.  The risk here is the so-called ``Volkswagen problem'' in which the diagnostic runs of the experiment are systematically different from when we run the algorithm of interest.

Another natural thought is to apply statistical tests to samples from the output distribution of a quantum supremacy experiment, to attempt to determine whether it is consistent with the desired distribution. A challenge for this approach is that many such tests require calculation of individual probabilities, which is assumed to be classically hard in the post-selection-based strategies.  Indeed a classical simulator with post-selection could simply guess a random output string and then post-select on it passing the verification.
This is an obstacle to simultaneously (a) basing our hardness on assuming $\PostBPP\neq\PostBQP$,  (b) efficiently verifying the output, and (c) having our verifier depend solely on the input and output to the quantum device.  However, examples exist of verification procedures that drop each of these.  Using factoring as a hard problem avoids using (a).  Using exponential classical time but polynomial quantum time\scite{google-circuit16,aaronson16} means dropping (b).  And we can drop (c) by generating a secret string $x$ and only providing some derived string $y=f(x)$ to the quantum computer.

In other words, we can attempt to find a task for which we know the answer in advance, but where we are able to hide it in such a way that it is likely hard to determine classically, while still being accessible to a quantum computer. In the case of IQP, Shepherd and Bremner\scite{SB09} proposed a task of this form based on, roughly speaking, hiding a bit string within a seemingly random linear code.   The verifier knows how the code has been scrambled but the quantum device (or classical simulator) sees only a version of the code from which the hidden string is not obvious.
A hidden correlation within the output probability distribution of the IQP circuit then encodes the identity of the hidden string. If the string is known in advance, one can test from a few samples whether the output distribution from the experiment corresponds to the correct string. However, if the hidden string is unknown, there is conjectured to be no structure in the code which would allow it to be determined by a classical algorithm. It is an interesting open problem to try to develop further challenges of this form which are based on more familiar cryptographic hardness assumptions.

One can also define efficient tests which rule out particular distributions as alternative explanations for the output of the supremacy experiment\scite{aaronson14a,spagnolo13,carolan14}. This can be reasonable if one has prior knowledge for suspecting that some distributions are likely, say if they correspond to the effects of decoherence.

The gold standard of verification would be direct certification of the quantum computation, where we check directly that the computation has worked, perhaps using information beyond merely the classical output of the experiment. In each case known so far, this approach requires more resources than performing the original computation. One example is a proposition by Hangleiter et al.\scite{HangleiterKSE16} that IQP computations could be verified by encoding them into the ground state of a local Hamiltonian based on a universality construction for adiabatic quantum computation\scite{gosset15}, and then checking that the state was prepared correctly using local measurements. This idea in fact works for any quantum computation. However, it loses the simplicity and noise-tolerance of the original IQP circuit, and requires one to believe that the experimenter has correctly implemented the local measurement operations. Another approach is the use of a distributed protocol to certify that a remote computer (or multiple computers) has performed an arbitrary quantum computation. This is exemplified by the model of blind quantum computation\scite{broadbent09}, which requires the verifier to be able to create single-qubit states and send them to the prover.  If we could make the verifier fully classical then such a verified quantum computation could be a way of experimentally confirming the computational predictions of quantum mechanics, analogous to the way that Bell experiments test the theory of entanglement\scite{AV13}.

In summary, all known verification techniques for quantum supremacy experiments have drawbacks: they are either inefficient in terms of the classical or quantum resources required, or assume that the behaviour of the experiment under test conditions corresponds to the real experiment, or make computational hardness assumptions which are not yet well understood. Developing verification techniques which avoid these issues is a pressing open question in quantum supremacy research.

\section*{Outlook}

In just a few years, quantum supremacy-type experiments have progressed from demonstrating boson sampling with 3 photons to a proposed implementation of random quantum circuits on 49 qubits.
Each of the diverse quantum supremacy proposals meets the five requirements we described at the start of this article, except for verification.
In parallel with experimental progress towards demonstrating quantum supremacy, improved classical simulation results have been proven for noisy and imperfect quantum supremacy experiments\scite{RahimiRC16,KalaiK14,bremner16b}. Thus the early stages of quantum supremacy experiments are likely to be characterised by an iterative process where proposed supremacy experiments are challenged by efficient classical simulations.  Nevertheless, given the speed of recent experimental developments, it seems plausible that quantum supremacy could be convincingly demonstrated in a matter of years.

There are many important open theoretical questions in the area of quantum supremacy.  The most pressing in our view is to develop a scheme that can be efficiently verified, by analogy with the way that the statistics for Bell tests can be easily checked.  Developing good classical simulations (or even attempting to and failing) would also help clarify the quantum/classical boundary.  The hardness assumptions could also be simplified and strengthened.  One ambitious goal in this direction would be to show that simulation with even low variational distance (cf.~\eq{trace-dist}) would imply the collapse of the $\PH$.
Theorists can and should also do more work to come to terms with two other models that appear to be non-universal for quantum computing but where we lack good classical simulations: finite-temperature adiabatic evolution with stoquastic Hamiltonians\scite{bravyi06,Dickson13,Nishimura16} (as used in the quantum annealers available from the company D-Wave) and analog quantum simulation\scite{cirac12,georgescu14}, for example of lattice models.   

We close by noting that supremacy is not a long-term goal but rather a necessary step in the development of quantum computers.  Eventually we expect that quantum computers will justify themselves by solving important problems which we do not know how to otherwise solve.  But in these early days of the field, the focus on quantum supremacy is a way to ensure that quantum computers solve clearly defined problems for which the classical competition can be well understood.

\section*{Acknowledgements}

AWH was funded by NSF grants CCF-1629809 and CCF-1452616. AM was supported by EPSRC Early Career Fellowship EP/L021005/1. AM would like to thank Mick Bremner and Dan Shepherd for introducing him to this topic, and for many debates concerning it over the years.

\section*{Selected references}
Some of the key readings on the topic are:
\bit
\item Terhal and DiVincenzo\cite{TD04}: This paper gave the first complexity-theoretic argument that a simple class of quantum circuits should be hard to simulate classically.
\item Aaronson and Arkhipov\cite{Aaronson13}: This seminal paper introduced the boson sampling problem.
\item Bremner, Jozsa and Shepherd\cite{BJS10}: This paper gave evidence that Instantaneous Quantum Polynomial-time (IQP) circuits are hard to simulate classically.
\item Boixo et al.\cite{google-circuit16}: This paper describes a proposal for a near-term quantum supremacy experiment.
\eit

\end{document}